# Far-infrared measurements of oxygen-doped polycrystalline $La_2CuO_{4.0315}$ superconductor under slow-cooled and fast-cooled conditions

Y. H. Kim[1+], H. H. Hsieh[2*], Z. Wu[2] and P. H. Hor[2++]

[1]Department of Physics, University of Cincinnati, Cincinnati, Ohio 45221-0011, U.S.A.
[2]Department of Physics and the Texas Center for Superconductivity
University of Houston, Houston, Texas 77204-5005, U.S.A.

We have studied the far-infrared (far-IR) charge dynamics of an equilibrated pure oxygen doped $La_2CuO_{4.0315}$ under slow-cooled and fast-cooled conditions. The superconducting transition temperature ($T_c$) for the slow-cooled and that for the fast-cooled processes were respectively found to be close to the two intrinsic $T_c$'s: One at 30 $K$ and the other at 15 $K$. Direct comparison with our previous results and other far-IR and Raman studies on single crystalline $La_{2-x}Sr_xCuO_4$, we conclude that the topology of the pristine electronic phases that are responsible for the two intrinsic $T_c$'s is holes arranged into two-dimensional (2D) square lattices.

PACS number: 74.25.Gz, 74.72.-h, 78.30.-j

*Present address: Department of Electrical Engineering, Chung Cheng Institute of Technology, National Defense University, Taoyuan, 335 Taiwan, R.O.C.

+ email address: kimy@ucmail.uc.edu
++ email address: phor@uh.edu

High $T_c$ superconductivity (HTS) in doped cuprates still remains enigmatic to the majority in the high $T_c$ research community. This is mainly due to the lack of a unified picture of the underlying electronic states responsible for the nano-scale electronic inhomogeneity[1] in the doping-induced charge carrier (hole) distribution in the $CuO_2$ planes. This hallmark of cuprate physics is now widely, though not universally[2], accepted as one of the natural consequences of the electronic phase separation. One possible candidate that we proposed[3] for the underlying pristine electronic phase (PEP) responsible for the nano-scale phase separation was the 2D electronic square lattices on which the itinerant holes "ride". Since the planar hole density $(P_{pl})$[4] is one of the most important controlling parameters of the HTS, it is plausible to assess that the generic electronic phase diagram, in which a superconducting dome exists in a narrow range between $P_{pl} \sim 0.06$ (or 0.11) and 0.25, arises from the presence of the PEP's and their interaction with the itinerant holes[5].

Hence, any theoretical model built on the premise where *all* the holes are itinerant in the $CuO_2$ planes cannot readily answer a question like "How could HTS emerge from such highly inhomogeneous electronic texture?" Consequently, theoretical interpretations to date have offered only a partial explanation for the data obtained by a particular experimental probe but fail to account for the rest observed through different probes, making numerous experimental data seem anomalous. Therefore, one may logically conclude that the missing key step in building the correct microscopic model is the pinning down of the origin of this ubiquitous electronic inhomogeneity and its role in the HTS.

At this point we need to bring out the two subtle but quite important experimental facts that have been overlooked so far. The first is the presence of the distinct onset $P_{pl}$'s for the HTS at $P_{pl} \sim 0.06$ (or 0.11) for the entire family of cuprate superconductors[4,5]. The second is the finding that only a small fraction of the holes are participating in the HTS. It was found that ~ 20% of the total holes are itinerant at optimal doping[6,7] and only ~ 1% in the underdoped $La_{2-x}Sr_xCuO_{4+\delta}$[3,8,9]. What is more fascinating is that the remaining 80% – 99% of the holes in the $CuO_2$ planes comprise the PEP's in the form of electronic lattices (EL's) as evidenced by the presence of the charge collective modes at finite frequencies and the accompanying single particle excitation gap at ~ 400 cm$^{-1}$, which is a direct proof that the EL's are self-organized and pinned[3,8,9].

In this work, we chose to study a polycrystalline $La_2CuO_{4+\delta}$ at $\delta = 0.0315$ which translates into $P_{pl} = 0.063$, right above the onset $P_{pl}$ for the HTS. Previous systematic studies of the preparations and the doping efficiency of pure oxygen-doped $La_2CuO_{4+\delta}$ and oxygen and strontium co-doped $La_{2-x}Sr_xCuO_{4+\delta}$ (CD-La214) polycrystalline samples found that delicate electrochemical oxidation performed at an elevated temperature ($T$) and long time post-annealing are required in order to achieve the thermodynamic equilibrium state[10,11]. These equilibrium samples have only two intrinsic superconducting transitions with $T_c = 15$ K and $T_c = 30$ K[10], which is independently confirmed later[12]. The two intrinsic $T_c = 15$ K and $T_c = 30$ K phases are not owing to the phase separation of the dopant oxygen atoms between the $CuO_2$ planes – these are the energetically favored electronic states as demonstrated in the studies of the electronic phase diagram of pure oxygen-doped $La_2CuO_{4+\delta}$ and CD-La214 polycrystalline samples under high pressure[13]. Although it was in a different context, the earlier studies of the hole cluster diffusion and oxygen diffusion in $La_2CuO_{4+\delta}$ as well as the electronic phase separation of the holes also reached the same conclusion.[14]

It was also found that depending on the cooling rate, one of the two superconducting phases could be selectively stabilized in $La_2CuO_{4+\delta}$ system. For instance, when the sample is cooled slowly down to 200 K above which the dopant oxygen atoms are highly mobile, the energetically most favored $T_c = 30\ K$ superconducting state results. However, the $T_c$ of the La$_2$CuO$_{4+\delta}$ with $P_{pl} \leq 1/16$, which always exhibits the $T_c = 30\ K$ superconductivity on slow cooling, can be brought to 15 K via quenching. Therefore, here we have a unique opportunity to test whether or not the notion of the mutually exclusive competing order is indeed operating behind the scene because, if they are mutually exclusive, the quenching process should induce EL's. Furthermore, our experiment eliminates the extrinsic disorder effects that may arise from two different samples at two different doping levels, guaranteeing that the observed changes are intrinsic.

For quenching, the sample mounted on the cold finger of a continuous liquid helium flow cryostat was submerged in liquid $N_2$. The quenching from room $T$ down to 77 K took less than 1 minute. Once the $T$ of the cold finger starts to drop below 77 K, the sample assembly was placed in the spectrometer for the reflectivity measurements. The entire procedure was done in a dry $N_2$ gas atmosphere. The slow cooling of the sample was performed under the normal operating condition of the cryostat by adjusting the temperature step by step at the cooling rate $\Delta T/\Delta t = -1\ K/\min$ down to $T = 160\ K$ and then rapidly cooled down to $T = 10\ K$. All the far-IR measurements were done on warming up of the sample from $T = 10\ K$.

The magnetic measurements were carried out in a magnetometer and the dc conductivity was measured by using a standard four-probe technique. Fig. 1a clearly shows superconductivity at $T_c \sim 20\ K$ upon quenching and at $T_c \sim 30\ K$ when slow-cooled. It is interesting to note that while both cases reached zero resistance, the resistivity of the quenched sample is "lower" than that of the slow-cooled one (see Figure 1b), which is a clear indication that the quenching process creates a larger fraction of "metallic" region corresponding to a $T_c \sim 20\ K$. The quenched and slow-cooled resistivity and magnetic susceptibility data were measured by following closely to the quenched and slow-cooled processes of the far-IR measurements.

The far-IR reflectivities of the quenched and of the slow-cooled sample at various $T$'s are displayed in Fig. 2. Overall reflectivity increases systematically with decreasing $T$. The reflectivity minimum at $\omega \sim 15$ cm$^{-1}$ and the peak at $\omega \sim 23$ cm$^{-1}$ are commonly present for both quenched and slow-cooled sample. Also a small peak develops at $\omega \sim 100$ cm$^{-1}$ in both cases with decreasing $T$. The intense mode at $\omega \sim 220$ cm$^{-1}$ (denoted as C in Figure 3) is the well-known c-axis breathing mode of the apical oxygen.

$\sigma_1(\omega)$ and $\varepsilon_1(\omega)$ calculated from the reflectivity data using a Kramers-Kronig transformation[3] are displayed in Fig. 3. The origin of the reflectivity minimum in both the slow-cooled and the quenched sample at $\omega \sim 15$ cm$^{-1}$ becomes clear; this is the characteristic of the plasma behavior of the free carriers in the system. The small value of $\omega_p$ suggests that the ratio of the free hole density to the hole mass is extremely small. In the $\sigma_1(\omega)$ plot for the quenched sample, the two peaks indicated as $G_1$ and P develop as $T$ is lowered below 300 $K$ and there also develop a broad structure between the $G_1$ and P modes.

Upon slow cooling, in addition to the $G_1$ and P modes observed in the quenched sample, two new modes ($G_2$ at $\omega \sim 43$ cm$^{-1}$ and $G_3$ at $\omega \sim 62$ cm$^{-1}$) emerge out of the broad structure and the $T_c = 30$ $K$ superconductivity results. Thus, one may conclude that the development of the $G_2$ (and $G_3$) mode must be related to the $T_c = 30$ $K$ superconductivity. In fact, the previous far-IR studies of the CD-La214[3] found that the development of the $G_2$ peak is essential for having the $T_c = 30$ $K$ superconductivity and it was suggested that this $G_2$ mode is the charge collective mode of the c(2x2) EL.[3,8] The single particle excitation gap $2\Delta \sim 400$ cm$^{-1}$ (50 meV) for both quenched and slow-cooled sample was estimated from the zero-crossing in $\varepsilon_1(\omega)$. However, this zero-crossing frequency must be higher than the actual zero-crossing of the gap because of the extra negative contributions to $\varepsilon_1(\omega)$ by the phonons. Notice that there is a discrepancy in the reflectivity obtained at 300 K under the quenched and the slow-cooled conditions (see Fig. 2). This is because it takes a substantial amount of time to "anneal away" the localized carriers in the EL sites even at room $T$ once the EL's are formed.

The development of the P mode upon doping along with the G modes deserves special attention because, although the P mode appears as a weak mode in the polycrystalline sample, this mode is dominating the in-plane $\sigma_1(\omega)$ of single crystalline $La_{2-x}Sr_xCuO_4$[9,15]. As shown in Fig. 4, direct comparison reveals that the P mode seen in the polycrystalline $La_2CuO_{4.0315}$ is a superposition of the two modes, one (Z mode) at ~ 100 cm$^{-1}$ which develops at low $T$ and the other ($X_1$ mode) at ~ 110 cm$^{-1}$ at low $T$ which is red-shifted from ~ 140 cm$^{-1}$ at room $T$. This Z mode must be related to the charge dynamics along the c-axis because it is absent in the in-plane $\sigma_1(\omega)$ of the single crystal sample. There also commonly presents a weaker but sharper mode at ~ 360 cm$^{-1}$ indicated as $X_2$.

These Z, $X_1$ and $X_2$ modes also appear in the Raman studies[16,17]. Lack of oxygen isotope effect on the Z mode and the $X_1$ mode in the Raman spectra[16] suggests that these are related to the motion of the La/Sr atom connected to the apical oxygen atom of each octahedron. The Z mode is highly polarized along the c-axis and the $X_1$ and $X_2$ modes are highly polarized along the $CuO_2$ plane. However the $X_2$ mode follows the mass harmonic law upon oxygen isotope substitution meaning that it is related to the apical oxygen vibration mode parallel to the $CuO_2$ plane[16]. It was also found that these Z, $X_1$ and $X_2$ modes developed only for the $P_{pl}$ range between 0.03 and 0.27, consistent with the far-IR observation. The Raman intensities of these modes mainly depended on the amount of doping and the local lattice distortions induced by doping, suggesting that the inversion symmetry breaking takes place due to the EL formation[17].

The asymmetric line shape of the $X_1$ mode is due to the Fano-type coupling[18], clearly indicating the strong phonon–EL interaction. The linewidth of $X_1$ mode in the polycrystalline sample (~ 8 cm$^{-1}$) is much narrower than that of the single crystalline sample (~ 40 cm$^{-1}$) at low $T$, suggesting that the degree of electronic disorder in the single crystalline sample is much more severe as pointed out in Ref.[9]. We point out that the ratio of the oscillator strength of the $X_1$ mode of the polycrystalline sample to that of the single crystalline sample is ~ 0.02. In other words, the spectral contribution of the $CuO_2$ plane to the reflectivity of the polycrystalline sample is only ~ 2%. Since the $X_1$ mode is highly in-plane polarized, it is expected to be more dominating the in-plane $\sigma_1(\omega)$ of the single crystalline sample as the angle of incidence approaches the normal incidence[9,15].

One-dimensional (1D) charge stripe order as the topology of the EL in the $La_{2-x}Sr_xCuO_{4+\delta}$ system was proposed based on the neutron study of $La_{1.6-x}Nd_{0.4}Sr_xCuO_4$ at $P_{pl}$ ~ 1/8.[19] Since superconductivity in $La_{2-x}Sr_xCuO_{4+\delta}$ is suppressed at $P_{pl}$ ~ 1/8, this was attributed to the charge stripe formation at $P_{pl}$ ~ 1/8. Then, according to this scenario, the static charge stripes must melt or become dynamic to bring about the HTS, which implies the diminishing of the oscillator strength of EL modes away from $P_{pl}$ ~ 1/8. This is not what has been seen in this work.

The previous electrochemical doping studies and the far-IR studies of $La_{2-x}Sr_xCuO_{4+\delta}$ led us to conclude that the PEP formed at $P_{pl}$ ~ 0.06 corresponds to a 2D EL of p(4x4) symmetry where $P_{pl}$ = 1/16 = 0.0625 which supports the $T_c$ = 15 K superconductivity. Upon further increasing the $P_{pl}$ the 2D EL of c(2x2) symmetry ($P_{pl}$ = 1/8 = 0.125) emerges and the $T_c$ = 30 K superconductivity results. Indeed, recent experimental evidences seem to agree with this observation that the intrinsic $T_c$'s of the HTS is tied to specific PEP's of $P_{pl}$ = 1/16 (= $1/4^2$), 1/8 (= $2/4^2$), 1/9 (= $1/3^2$), 3/16 (= $3/4^2$), and 1/4 (= $4/4^2$), which naturally leads to the 2D EL's[20]. However, the far-IR evidence for the 2D nature of the EL was deduced from the line shape of the single particle excitation peak that lacks the characteristics of the 1D EL[3].

In principle it is possible to test the dimensionality of the EL's by measuring the far-IR in-plane anisotropy. This has not been feasible for our polycrystalline samples and even for single crystalline samples because of the twinning problem. Fortunately, Padilla *et al*[21] reported detailed far-IR studies of the de-twinned $La_{2-x}Sr_xCuO_4$ single crystals as a function of doping for two polarizations, one along the *a*-axis and the other along the *b*-axis of the $CuO_2$ plane. They observed the broad peak at ~ 110 $cm^{-1}$ in both directions[22]. This broad peak at ~ 110 $cm^{-1}$ is the charge-induced $X_1$ mode described in this paper and its strength is already dominating even at ~ 4% doping. Therefore, there is no intrinsic anisotropy in the in-plane far-IR spectra and the EL is indeed 2D in nature.

In light of the two intrinsic superconducting phases ($T_c$ = 15 K and $T_c$ = 30 K) observed in the $La_2CuO_{4+\delta}$ in thermal equilibrium, one may conclude that the $T_c$ = 20 K superconducting state observed in the quenched sample is an admixture of the $T_c$ = 15 K phase and of the $T_c$ = 30 K phase. Hence the $G_2$ mode (and $G_3$ mode) is also present in the quenched sample but they appear as a broad bump. Upon slow-cooling, the $G_2$ mode dominates and the $T_c$ = 30 K state is obtained. However we believe that the long-range order of the $P_{pl}$ = 3/16 EL of the $G_3$ mode, which should correspond to the $T_c$ = 45 K phase, was not fully developed to support the $T_c$ = 45 K superconductivity[23] and must have been present in the form of small disconnected patches in the $CuO_2$ planes.

In conclusion, we have verified the 2D nature of the EL through detailed analysis of the charge-induced infrared modes. Our observation is contrary not only to the notion of the mutually exclusive competing order but also to the 1D static/dynamic charge stripe order as the topology of the EL. In this work, we have identified for the first time the physical origin of the mysterious intense structure at ~ 110 cm$^{-1}$ that has been misunderstood as a charge stripe related structure.[24] We have demonstrated that the presence of 2D EL is essential for the appearance of the HTS. The severe electronic disorder in the single crystalline sample suggests mixing of the disordered EL's of various symmetries[9]. We anticipate, therefore, inhomogeneous local gaps in the STM experiments since the so-called pseudo gap is the single particle excitation gap of an EL[25]. The STM observation of the co-existence of 100% coverage of the nano-scale inhomogeneous local gaps with HTS[26] and the non-competing nature of the pseudo gap against HTS[27] is consistent with our composite picture of a small amount of free carriers moving on the mixed 2D EL's that form the condensate below $T_c$[28].

We thank Z. G. Li for preparing the equilibrium electrochemically charged La$_2$CuO$_{4.0315}$ sample for us. Z. Wu, Z. G. Li, H. H. Hsieh, and P. H. Hor are supported by the State of Texas through the Texas Center for Superconductivity at the University of Houston.

**Figure Captions:**

**Figure 1.** (a) The magnetic susceptibility measured under field-cooled at 5 Gauss and (b) the resistivity of the quenched and slow-cooled $La_2CuO_{4+0.00315}$. The data is always collected during warming from 5 $K$.

**Figure 2.** Far-IR reflectivity of the quenched and slow-cooled $La_2CuO_{4+0.0315}$. From bottom to top: $T$ = 300 $K$, 270 $K$, 250 $K$, 220 $K$, 200 $K$, 190 $K$, 180 $K$, 170 $K$, 160 $K$, 130 $K$, 110 $K$, 100 $K$, 70 $K$, 50 $K$, 40 $K$, 35 $K$, 34 $K$, 33 $K$, 32 $K$, 31 $K$, 30 $K$, 29 $K$, 28 $K$, 27 $K$, 26 $K$, 25 $K$, 23 $K$, 22 $K$, 21 $K$, 20 $K$, 19 $K$, 18 $K$, 17 $K$, 16 $K$, 15 $K$, 12 $K$, and 10 $K$. The black lines are for the superconducting state.

**Figure 3.** Temperature of $\sigma_1(\omega)$ and $\varepsilon_1(\omega)$. Black lines indicate the superconducting state. The $T$ in the lower panels decreases from top to bottom.

**Figure 4.** Direct comparison of the EL-induced modes of $La_2CuO_{4+0.0315}$ with those of the $La_{1.93}Sr_{0.07}CuO_4$ single crystal ($T_c$ ~ 20 $K$). From bottom to top: T = 300 $K$, 200 $K$, 100 $K$, 50 $K$, 30 $K$, 24 $K$, 16 $K$, and 8 $K$ for $La_{1.93}Sr_{0.07}CuO_4$; T = 300 $K$, 250 $K$, 200 $K$, 180 $K$, 160 $K$, 100 $K$, 70 $K$, 40 $K$, 30 $K$, and 10 $K$ for $La_2CuO_{4+0315}$. Note the log-scale in frequency.

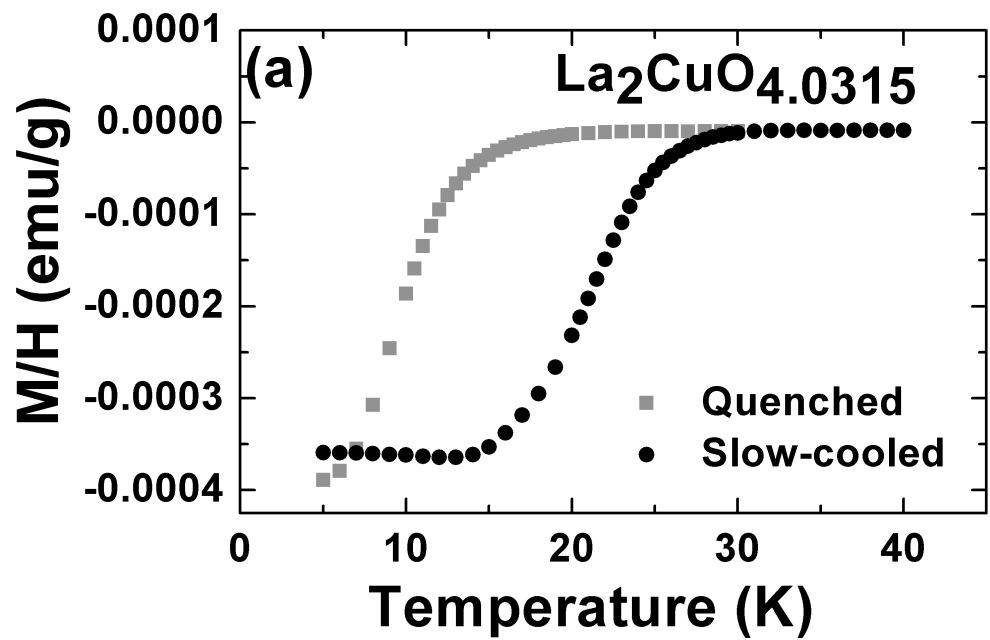
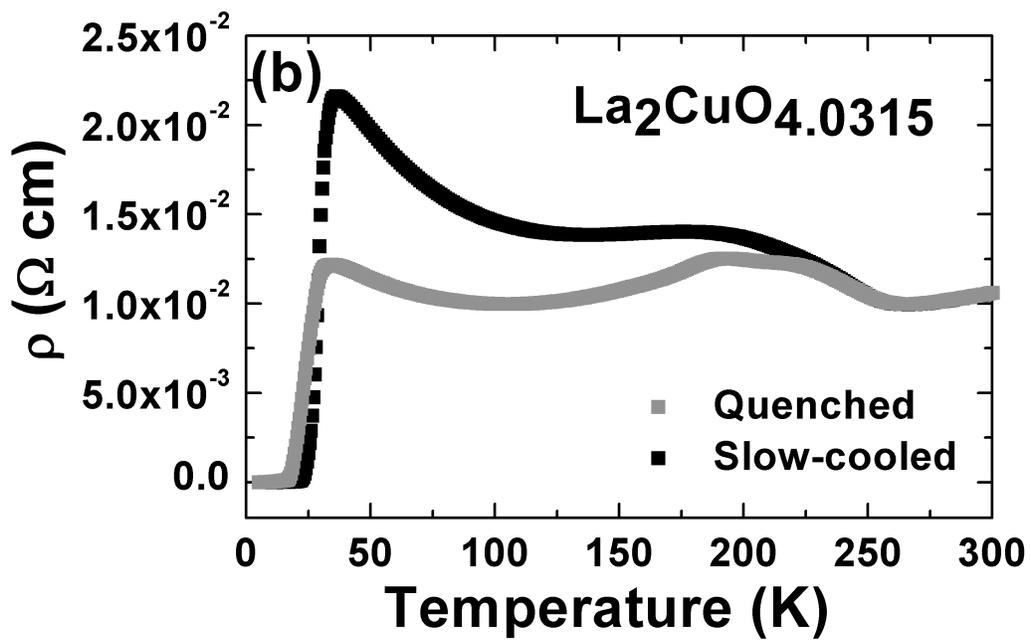

Figure 1

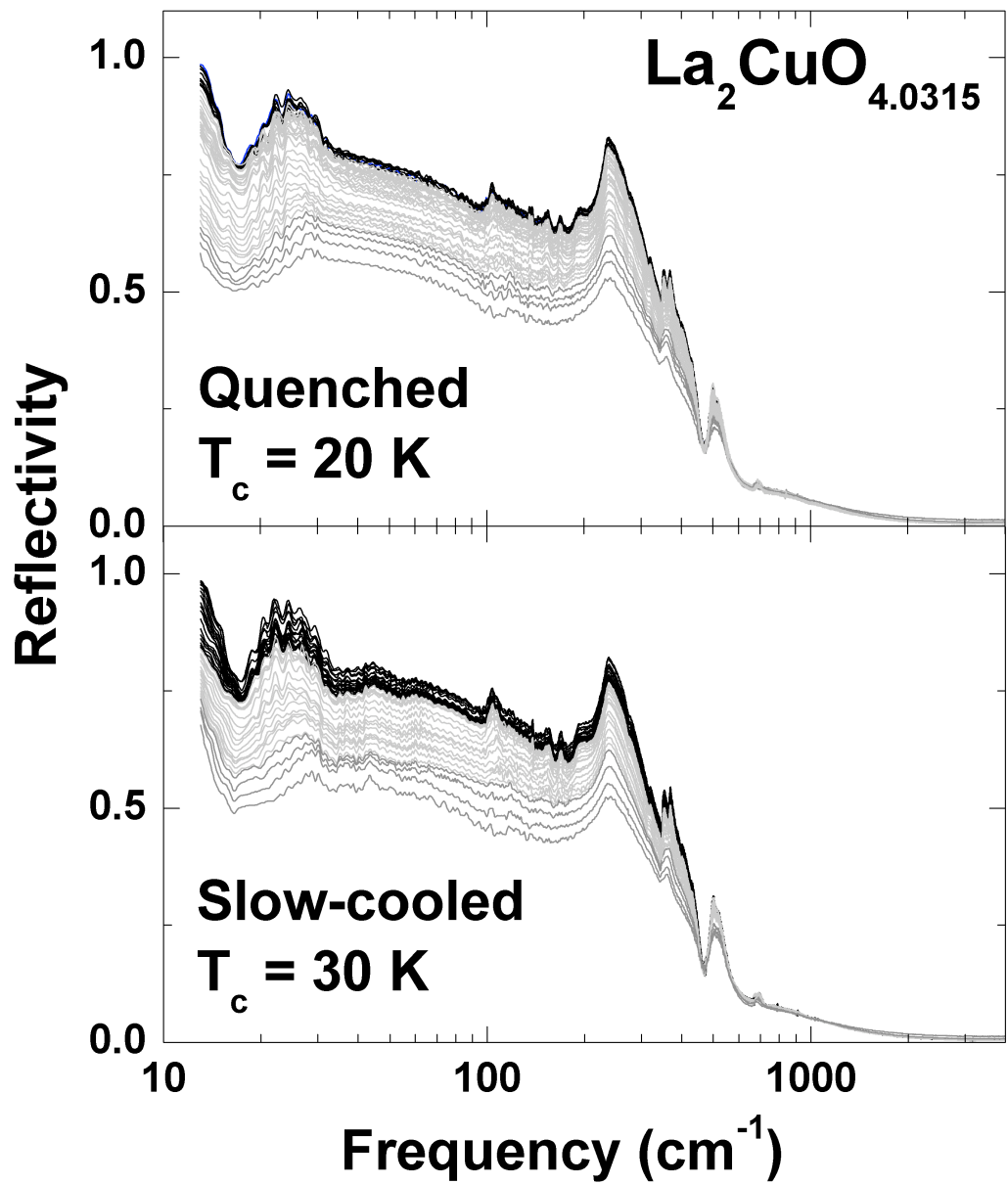

**Figure 2**

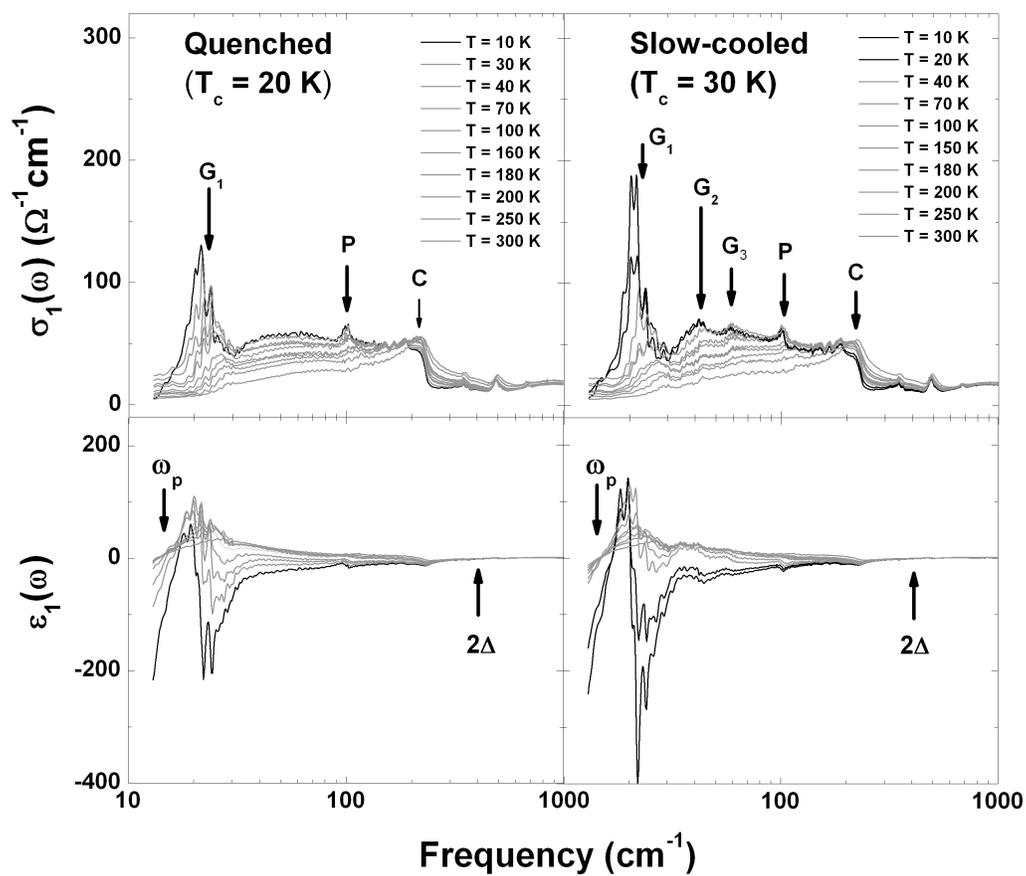

Figure 3

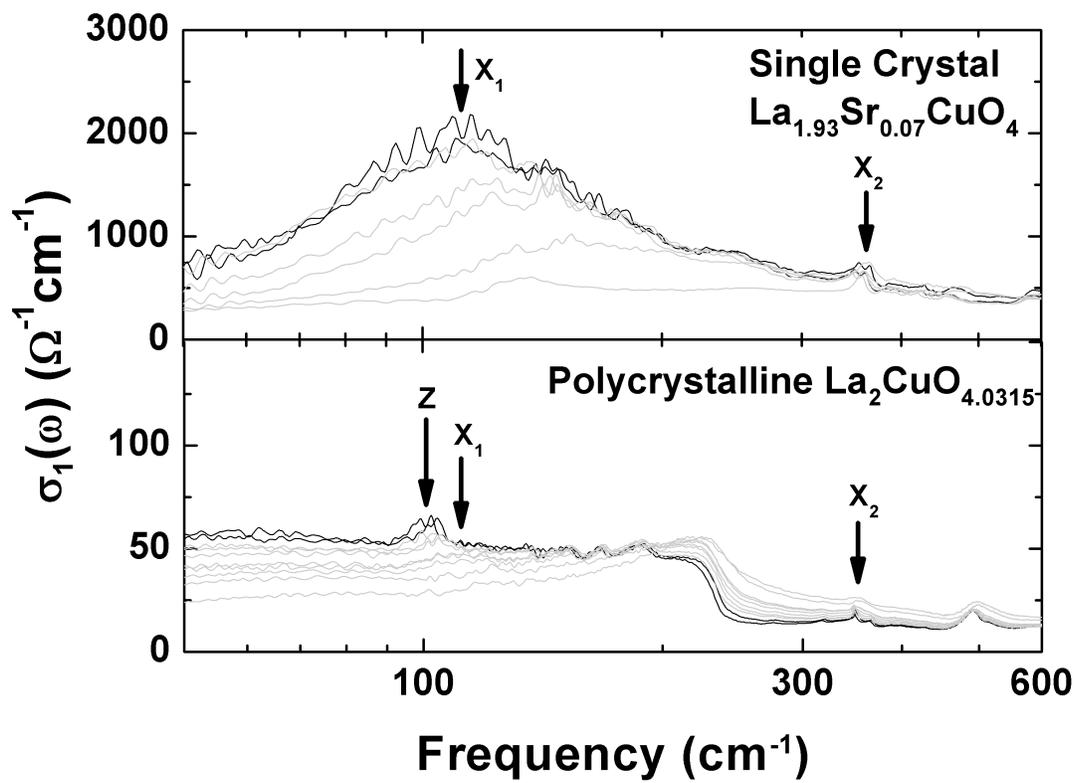

Figure 4